\documentclass[a4paper,11pt]{article}
\usepackage{pos}

\title{Extraction of Information from Polarized Deep Exclusive Scattering with Machine Learning}

\author{Simonetta Liuti}

\affiliation{Physics Department, Univeristy of Virginia, Charlottesville VA 22904, USA.}

\emailAdd{sl4y@virginia.edu}

\abstract{A framework defining benchmarks for the analysis of polarized exclusive scattering cross sections is proposed that uses physics symmetry constraints as well as lattice QCD predictions. These constraints are built into machine learning (ML) algorithms. Both physics driven and ML based benchmarks are applied to a wide range of deeply virtual exclusive processes through explainable ML techniques with controllable uncertainties. The observables, namely the Compton Form Factors (CFFs) which are convolutions of Generalized Parton Distributions (GPDs), are extracted using methods such as the random targets method to evaluate the separate contribution of the aleatoric and epistemic uncertainties in exclusive scattering analyses.}

\FullConference{25th International Spin Physics Symposium (SPIN 2023)\\
 24-29 September 2023\\
 Durham, NC, USA\\}
 
\begin{document}
\maketitle

\section{Introduction}
\label{sec:intro}
Understanding the dynamical parton substructure of the nucleon in both momentum and coordinate space is essential to pinning down the working of angular momentum,  a fundamental goal of the physics program at both Jefferson Lab and at the upcoming electron ion collider (EIC) \cite{AbdulKhalek:2021gbh}. Key experiments are deeply virtual exclusive scattering  (DVES) processes where either a high momentum photon ($\gamma$) or a meson ($M$), is detected along with the recoiling proton. One can then access the spatial distributions of quarks and gluons through Fourier transformation of the processes matrix elements in the momentum transfer between the initial and scattered proton. 

\begin{figure}[h]
    \centering
    \includegraphics[width=8cm]{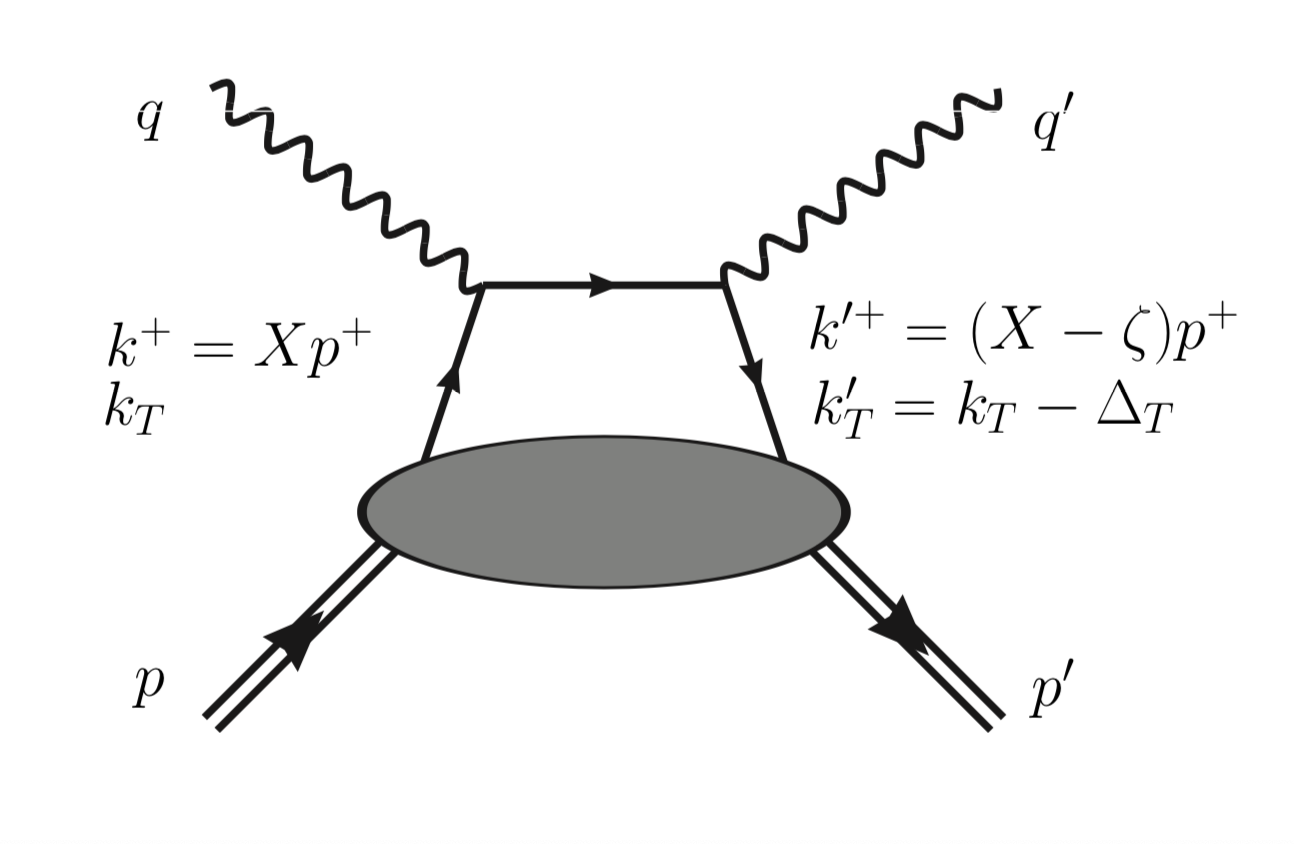}
    \caption{GPD Feynman diagram at tree level, illustrating the kinematics for DVES.}
    \label{fig:feynman}
\end{figure}

Assuming the validity of quantum chromodynamics (QCD) factorization theorems \cite{Ji:1997nk,Ji:1998xh} one can single out the correlation function for these processes, which are parametrized in terms of generalized parton distributions (GPDs). GPDs depend on the set of kinematic invariants $(Q^2, x_{Bj}, t, x)$, where $Q^2$, is the exchanged virtual photon four-momentum squared; $x_{Bj}$ is proportional to the so-called skewness parameter $\xi$ measuring the momentum transfer along the light-cone; the Mandelstam invariant $t$, gives the proton four-momentum transfer squared; $x$, the longitudinal momentum fraction carried by the struck parton (see Figure \ref{fig:feynman} and  Refs.\cite{Diehl:2003ny,Belitsky:2005qn,Kumericki:2016ehc} for a review of the subject). 
Similarly to the electromagnetic and weak form factors in elastic $ep$ scattering, the CFFs parametrize the amplitude of the DVES process, and they enter the cross section in bilinear/quadratic forms.
\footnote{At variance with elastic scattering, due to the extra degree of freedom given by the emitted photon or meson, the amplitude for DVES is a complex quantity}
GPDs -- the structure functions of the correlation function --  enter the CFFs -- the experimental observables -- only through convolutions over $x$, with Wilson coefficient functions which have been determined in perturbative QCD (PQCD) up to NLO in Ref.\cite{Kumericki:2019mgk} and NNLO in Refs.\cite{Braun:2022bpn}. The kinematic variable $x$, therefore, appears as an internal loop variable and is not directly observable. As a consequence, all information on the longitidinal momentum distributions of partons cannot be directly measured  

In this talk we presented a program for determining the spatial structure of the nucleon and angular momentum from experiment where information from QCD phenomenology and lattice QCD {\it instructs} machine learning (ML) methods. Towards this goal we joined efforts in the EXCLusives with AI and Machine learning (EXCLAIM) collaboration \cite{Exclaim}. EXCLAIM is developing, on one side, a solid statistical approach to the multidimensional inverse problems associated with the extraction of spatial structure from data. Using the latter as a backdrop, physics informed networks are designed that include theory constraints in deep learning models. In this approach, ML is not treated as a set of “black boxes” whose working is not fully controllable: our goal is to open the boxes using tools from information theory and quantum information theory to learn about their relevant mechanisms within a theoretical physics perspective and to interpret the ML algorithms which are necessary to extract information from data


%
%

\section{Benchmarks for Global Analysis of Deeply Virtual Exclusive Experiments}
The starting point of our study is to propose a possible set of phenomenology and ML benchmarks required for a precise determination of CFFs. Benchmarks are a necessary step which is needed to up the stage for the extraction of GPDs and related information on hadronic 3D structure from the CFF convolutions \cite{Almaeen:2022imx}. 
ML algorithms have already been used extensively to study high dimensional, ``raw", experimental data. Nevertheless, their use in theory and phenomenology is still rather new (see {\it e.g.} discussion in \cite{Carleo:2019ptp}). 
ML methods provide an alternative to parametric fitting procedures, including earlier ANN-based ones \cite{Ethier:2020way}, in which a functional form could bias the results when generalized to regions far outside of the data sets. 

\begin{figure}
    \centering
    \includegraphics[width=15cm]{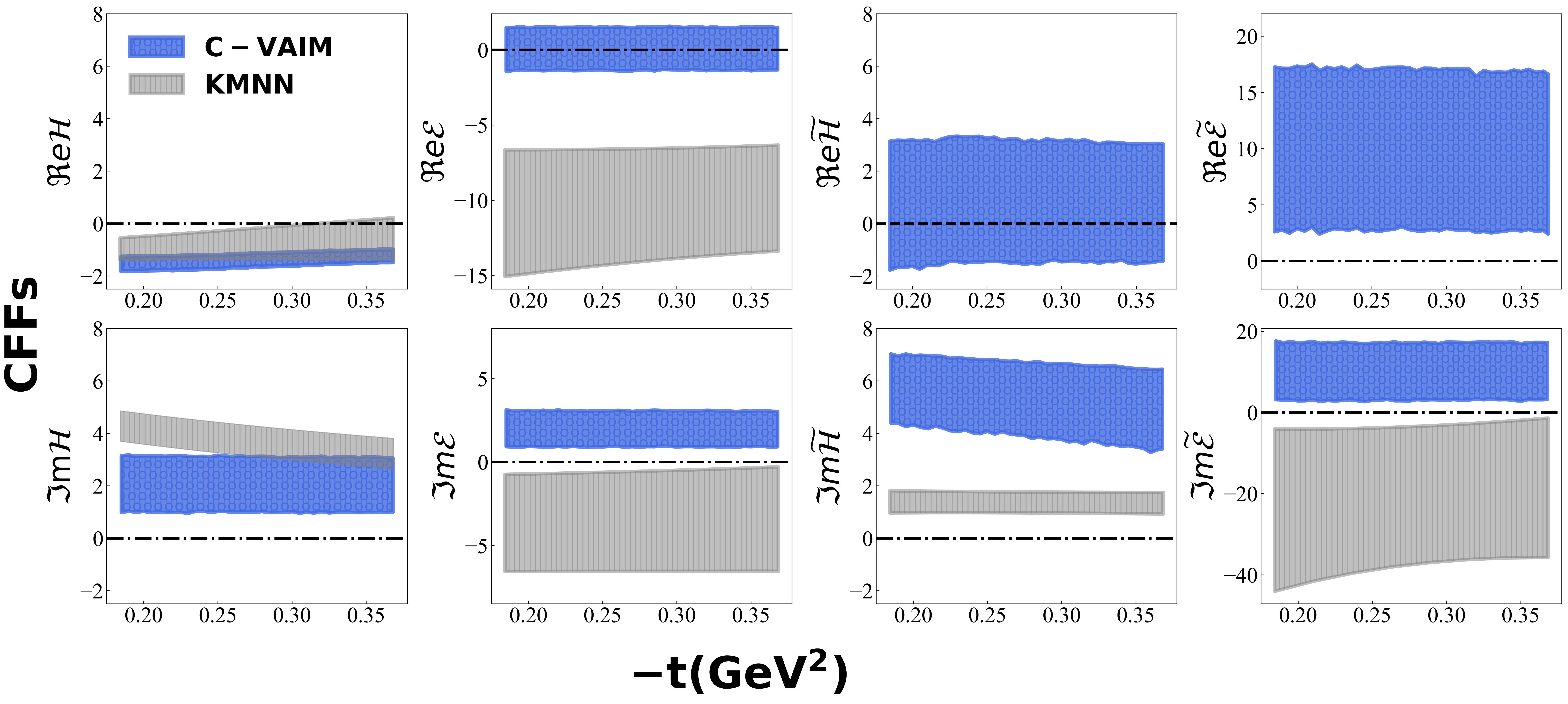}
    \caption{CFFs extracted from the VAIM approach of Ref.\cite{Almaeen:2024guo} and compared with results using an ANN approach from Ref.\cite{Cuic:2020iwt}.}
    \label{fig:CFF_VAIM}
\end{figure}

It should be noticed that some subsets of the benchmarks have already been addressed, in particular, through the efforts for the precision extraction of Parton Distribution Functions (PDFs) from a global analysis of high energy inclusive scattering data spearheaded by NNPDF (\cite{NNPDF:2021uiq} and references therein), CTEQ \cite{Hou:2019efy}, and MHST (formerly MSTW)  \cite{Watt:2012tq,Bailey:2020ooq} (for reviews see \cite{Amoroso:2022eow,Begel:2022kwp} and references therein).
In particular, NNPDF adopted machine learning techniques and new hyperparameter optimization methods that 
impact considerably the PDF uncertainties, bringing them close to the percent precision level. 
The purpose of defining the benchmarks is to provide a common set of similar rules that the community can come together to agree on. 
Using the Hessian method, the CTEQ collaboration has performed impact studies of the datasets from the LHC incorporated in PDF fitting on calculating benchmark processes such as the Higgs boson cross section \cite{Hou:2019efy}. The MHST collaboration combines data from the LHC and HERA to determine PDFs and uncertainties \cite{Bailey:2020ooq}.
In the exclusive processes sector, several groups have already been proposing extractions of CFFs using various approaches differing both in their numerical and analytic components, {\it e.g.} using different formalism/approximations for the cross section, and/or different data sets and kinematic ranges   
with important distinctions that make GPDs fundamentally different thus requiring a completely new, dedicated ``ground-up" approach \cite{Cuic:2020iwt,moutarde2019unbiased}. 
Most important, GPDs enter the exclusive cross sections at the amplitude level, similarly to the elastic proton form factors,  while PDFs define directly the inclusive cross section, therefore their extraction from experimental data requires solving a non trivial inverse problem. 
The framework presented here utilizes physics-informed ML models with architectures that are designed to  
satisfy some of the physics constraints from the theory by limiting the predictions to only those allowed by the theoretical input, resulting in less modeling error, a homogeneous treatment of data point, and faster training. A major advantage is also in the improved generalization, that helps us provide a more sound guidance for extracting more accurate results from experimental measurements. 
Theoretical physics ideas can be introduced in deep learning models as ``hard" constraints
by building them into the architecture of the network itself {\it e.g.} imposing  network invertibility, by appropriately choosing the activation functions, and by defining customized neural network layers.
Another way of introducing physics constraints in the   ``Soft" constraints are imposed by adding an additional term to the loss function that can be learned to approximately minimize. In other words, the effect of this term is to generate physics weighted parameters.
In Figure \ref{fig:CFF_VAIM} we show results from our analysis using a Conditional-Variational Autoencoder Inverse Mapper (C-VAIM) with constraints from: 1) Symmetries in the cross section structure; 2) Lorentz invariance; 3) Positivity;  4) Forward kinematic limit, defined by $\xi, t \rightarrow 0$, to PDFs, when applicable. An additional constraint from  $\Re e$ - $\Im m$ connection of CFFs through dispersion relations  with proper consideration of threshold effects was not applied in this case.

\section{Theory of DVES}
To optimize the extraction of information from data, one has to have a full and detailed understanding of the cross section for DVES processes \cite{Kriesten:2020wcx} (see also Ref.\cite{Qiu:2022pla}). The DVCS cross section was restructured in its BH, DVCS  and  BH-DVCS interference contributions using the helicity amplitudes based definitions of Refs.\cite{Arens:1996xw,Diehl:2005pc}. The resulting formalism organizes the cross section in a similar way to the one for elastic $ep$  scattering 
in terms of the nucleon charge, magnetic, and axial current contributions (the latter is reviewed in Refs.\cite{Perdrisat:2006hj,Gao:2003ag}).
This allows us to give a physical interpretation of the various terms, including the specific dependence on the angle $\phi$ between the lepton and hadron planes. The treatment of this phase factor for the virtual photon polarization vectors, explained in Refs.\cite{Kriesten:2019jep}, represents the most important difference with previous approaches, affecting the QCD separation of twist-two and twist-three terms.
\begin{figure}
    \centering
    \includegraphics[width=7cm]{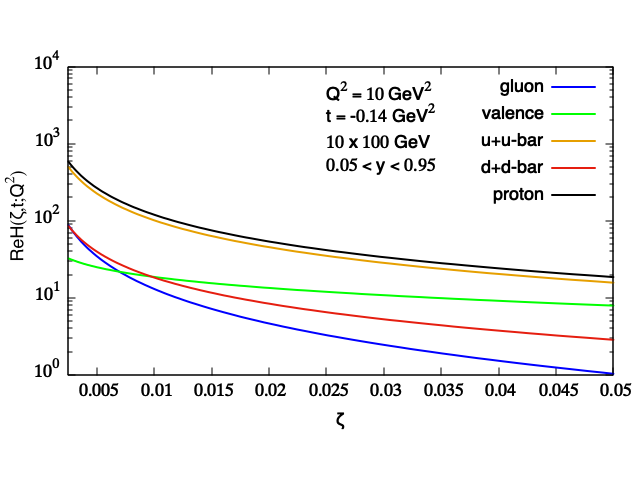}
    \includegraphics[width=7cm]{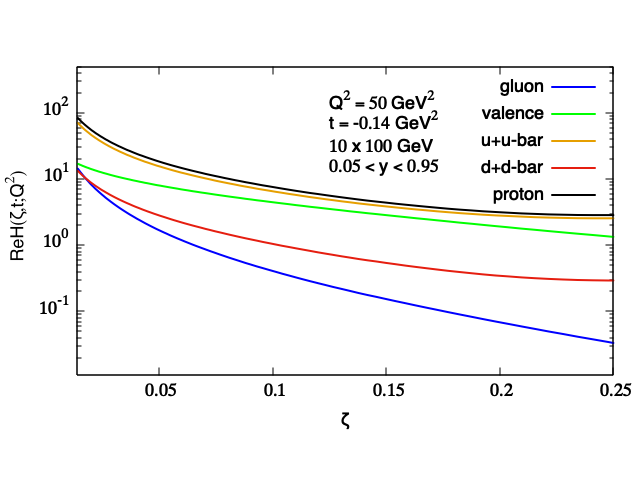}
    \caption{Contributions to the proton CFF as a function of $x_{Bj} = \zeta$ at a fixed kinematics $t = -0.14$ GeV$^{2}$. (\textit{Left}) The proton CFF is separated into gluon, valence, $u+\bar{u}$, and $d+\bar{d}$ components, where all components are scaled by their charges $e_{u}^{2} = 4/9$ and $e_{d}^{2} = 1/9$ at a fixed $Q^{2} = 10$ GeV$^{2}$. (\textit{Right}) The same contributions as the plot on the left for $Q^{2} = 50$ GeV$^{2}$. All predictions are calculated to ensure that $0.05 < y < 0.95$.}
    \label{fig:cff_vxbj_eic}
\end{figure}
The cross section for $e(k) + p(p) \rightarrow e'(k') + p'(p') + \gamma (q')$, on unpolarized proton is derived from a
coherent superposition of the DVCS and Bethe-Heitler (BH) amplitudes is written as (we refer the reader to Refs.\cite{Kriesten:2019jep,Kriesten:2020apm} for details).
The amplitude for the DVCS process shown in Figure \ref{fig:feynman} reads,
\begin{eqnarray}
\label{eq:TDVCScov}
T_{DVCS} & = &  e^3 j_{DVCS}^\mu \frac{\tilde{g}_{\mu \nu}}{q^2} J_{DVCS}^{\nu} 
\end{eqnarray}
with the lepton and hadron currents being respectively given by,
\begin{subequations}
\begin{eqnarray}
j_{DVCS}^\mu(q) &= & \overline{u}(k',h)\gamma^\mu u(k,h) 
\\
J_{DVCS}^{\nu}(q,q') & = & {\cal W}^{\mu \nu}(p,p') \left(\varepsilon^{\Lambda_{\gamma'}} _\mu(q') \right)^* \, .
\end{eqnarray}
\end{subequations}
where $q=k-k'$,
 $\varepsilon^{\Lambda_{\gamma'}}_{ \mu}(q')$ is the polarization vector of the outgoing photon, $\gamma'$.
${\cal W}^{\mu \nu}$ is the DVCS hadronic tensor \cite{Belitsky:2001ns}, parametrized in terms of the GPD correlation functions of twist-two ($W^{\gamma^+}, W^{\gamma^+\gamma_5}$), and twist-three ($W^{\gamma^i}, W^{\gamma^i\gamma_5}$) \cite{Meissner:2009ww},
 \begin{eqnarray}
 \label{eq:DVCStensor}
{\cal W}^{\mu \nu}  =  \frac{1}{2} \Big[ \Big(- g_T^{\mu\nu} W^{\gamma^+} + i \varepsilon_T^{\mu\nu} W^{\gamma^+\gamma_5} \Big)
 + \frac{2M x_{Bj}}{\sqrt{Q^2}} (q+ 4 \xi P)^\mu \Big(- g_T^{\nu i} W^{\gamma^i} + i \varepsilon_T^{\nu i} W^{\gamma^i\gamma_5} \Big) \Big] .
  \end{eqnarray}  
Using the photon projection operator, $\tilde{g}_{\mu \nu}$,  \cite{Dmitrasinovic:1989bf,Boffi:1993gs},
\begin{eqnarray}
\tilde{g}_{\mu \nu} = \sum_{\Lambda_{\gamma^*}} (-1)^{\Lambda_{\gamma^*}} \left[ \varepsilon_\mu^{\Lambda_{\gamma^*}}(q) \right]^* \varepsilon_\nu^{\Lambda_{\gamma^*}} (q)  ,
\end{eqnarray}
we can project out the contributions from the transverse ($\Lambda_{\gamma^*}= \pm 1 \equiv T$), and longitudinal ($\Lambda_{\gamma^*}=0 \equiv L$) polarized virtual photon, $\gamma^*(q)$. From the structure of Eq.\eqref{eq:DVCStensor}, analgously to deep inelastic scattering, one can immediately associate the photon transverse polarization to twist-two GPDs and the longitudinal polarization to twist-three GPDs.
Inserting the expansion in Eq.(\ref{eq:TDVCScov}) we obtain the following invariant expression,
\begin{eqnarray}
T_{DVCS} & = &  \frac{e^3}{q^2}  \left(j_{DVCS}^\mu \varepsilon_\mu^{\Lambda_{\gamma^*}} \right)^* \left( J_{DVCS}^{\nu} \varepsilon_\nu ^{\Lambda_{\gamma^*}} \right) ,
\end{eqnarray}
where the photon polarization vector contracted with the hadron current is evaluated in the hadron scattering plane, and it is therefore rotated by a phase,
\begin{equation}
    \varepsilon_\mu^{\Lambda_\gamma^*} (hadron) = e^{-i \Lambda_\gamma^* \phi} \, \varepsilon_\mu^{\Lambda_\gamma^*} (lepton) 
\end{equation}
The phase $\phi$  determines the structure of DVCS contribution to the cross section,
\begin{eqnarray}
\label{eq:TDVCS}
\mid T_{DVCS} \mid^2  =  F_{T} + \epsilon F_{L} + \sqrt{\epsilon (1-\epsilon)} F_{LT} \cos \phi  
+ \epsilon F_{TT} \cos 2 \phi \nonumber \\
\end{eqnarray}
where $\epsilon \equiv \epsilon_{DVCS}$, the ratio of longitudinal to transverse virtual photon polarization is given by,
\begin{eqnarray}
  \epsilon_{DVCS} & = &  \frac{1-y-\frac{1}{4}y^2\gamma^2}{1-y+\frac{1}{2}y^2 +\frac{1}{4}y^2\gamma^2} \nonumber \\
&  = & \frac{\sum_h \mid j^\mu_{DVCS} (\varepsilon_\mu ^{0}(q))^* \mid^2} 
{\sum_h \, \mid j^\mu_{DVCS} (\varepsilon_\mu ^{+1}(q))^* + j^\mu_{DVCS} (\varepsilon_\mu ^{-1}(q))^* \mid^2 } \nonumber \\
\label{eq:eps_DVCS}
 \end{eqnarray}
with $y=(kq)/(pq)$, $\gamma^2 = 4M^2x_{Bj}^2/Q^2$. The subscripts, $L,T$, refer to the virtual photon polarization for the matrix element modulus squared with same polarization for the initial and final virtual photons, while the terms $F_{LT}$ and $F_{TT}$ are transition elements from $L\rightarrow T$, and $T=\pm 1 \rightarrow T=\mp 1$, respectively. 
\footnote{In the context of Refs.\cite{Kriesten:2019jep,Kriesten:2020apm}, which addressed all beam-target polarization configurations, we used the notation: $F_{UU,(T,L)}$ for $F_{T,L} $.}

As shown in detail in Ref.\cite{Kriesten:2019jep}, using the GPD notation of \cite{Meissner:2009ww}, $F_T$ is described in terms of products of twist-two CFFs,
\[{\cal F}^* {\cal G} + {\cal G}^* {\cal F} \quad\quad {\rm with} \quad\quad {\cal F}, {\cal G} = {\cal H, E}, \widetilde{\cal H}, \widetilde{\cal E} ;\]
$F_{L}$ is given by the product of two twist-three CFFs, namely, 
\[({\cal F}^{(3)})^* {\cal G}^{(3)} + ({\cal G}^{(3)})^* {\cal F}^{(3)},  \]
with,
\[{\cal F}^{(3)}, {\cal G}^{(3)} = {\cal H}_{2T}, {\cal E}_{2T}, \widetilde{\cal H}_{2T}, \widetilde{\cal E}_{2T}, {\cal H'}_{2T}, {\cal  E'}_{2T}, \widetilde{\cal H'}_{2T}, \widetilde{\cal E}'_{2T}; \]
and 
$F_{LT}$ is given by the product of twist-two and twist-three CFFs, 
\[ {\cal F}^* {\cal F}^{(3)} + ({\cal F}^{(3)})^* {\cal F}    \]
Finally, $F_{TT}$, corresponds to a helicity flip of two units which can be only described in terms of transversity gluon degrees of freedom. We disregard this term in the present analysis since it is  suppressed by a factor $\alpha_S$ \cite{Belitsky:2001ns,Kriesten:2019jep}. 
The twist-two CFF, $\Re e {\cal H}$ is presented in Figure \ref{fig:cff_vxbj_eic}, for typical EIC kinematics, using the parametrization from Ref.\cite{Kriesten:2021sqc} where the contribution of valence, sea quarks and gluons is highlighted. 
Performing an L/T separation will give us access to the GPD twist three terms (for a full description of the GPD content of these terms see Refs.\cite{Kriesten:2019jep,Kriesten:2020apm}). 

The interference contribution to the cross section takes the form,
\begin{eqnarray}
{\cal I} = T_{DVCS}^* T_{BH} +  T_{BH}^* T_{DVCS} 
\end{eqnarray}
where $T_{DVCS}$ is given in Eq.(\ref{eq:TDVCScov}) and the BH term is described in \cite{Kriesten:2019jep}.  
One can work out the leptonic and hadronic contributions to these terms, which are, respectively given by
\begin{eqnarray}
j_\mu^{BH}(\Delta) (j_\nu^{DVCS}(q))^* & = & L_{\mu\rho} \, (\varepsilon^\rho(q'))^{ *} \, \, \left( \bar{u}(k') \gamma_\nu u(k)  \right)^* \nonumber  \\ \\
J_\mu^{BH}(\Delta) (J_\nu^{DVCS}(q))^* & = & \overline{U}(p') \Gamma_\mu U(p)  \, \, \left( {\cal W}_{\nu\rho} \varepsilon^\rho(q') \right)^* \nonumber \\  
\end{eqnarray}
with analogous expressions for the complex conjugates. The hadronic tensor is defined as,
\begin{eqnarray}
W^{\mathcal{I}}_{\mu \nu}   =  J_\mu^{BH}(\Delta) (J_\nu^{DVCS}(q))^* + (J_\mu^{BH}(\Delta))^{*} J_\nu^{DVCS}(q) \nonumber \\
\end{eqnarray}
In this case, the lepton and hadron tensors correspond to a mixed virtual photon representation 
with the DVCS photon, $\varepsilon^{\Lambda_\gamma^*}(q)$, and BH photon, $\varepsilon(\Delta)$. 
\footnote{Notice that the equations feature explicitly the outgoing real photon polarization, $\varepsilon^{\Lambda\gamma'}(q')$, where we omit the dependence on the polarization index since this is summed over.}
The leading, twist-two, contribution to the  hadronic tensor is obtained using the first line of the definition of the DVCS tensor ${\cal W}_{\nu\rho}$, defined in Eq.\eqref{eq:DVCStensor}, where $\varepsilon^{\Lambda_\gamma^*}(q)$ is transversely polarized,
\begin{eqnarray}
\label{eq:Wmu_DVCStw2}
W^{\mathcal{I}, \, {\rm tw 2}}_{\mu \nu} &=& P_{\mu} \varepsilon^*_\nu(q') \Big[F_{1}\mathcal{H} + \tau F_{2}\mathcal{E} \Big] + \Big[\frac{\xi}{2}\Delta_{\mu} 
 +  \frac{t}{4P^{+}}g_{\mu -} \Big] \varepsilon^*_\nu(q')  (F_{1}+F_{2})(\mathcal{H}+\mathcal{E}) \nonumber \\ 
&+ & \frac{1}{2P^{+}} \left(\epsilon_{\alpha \mu \beta -} {P}^\alpha \Delta^\beta \right) 
(\epsilon_T)_{\delta \nu} \left( \varepsilon^\delta (q') \right)^* \, \widetilde{\mathcal{H}}  (F_{1}+F_{2}) . 
\end{eqnarray}
The BH-DVCS interference contribution is expressed in terms of linear combinations of products of CFFs  elastic form factors, $F_1$ and $F_2$, with the coefficients, $A_{UU}^{\cal I}$, $B_{UU}^{\cal I}$, $C_{UU}^{\cal I}$, which are functions of $(Q^2, x_{Bj}, t, y, \phi)$ \cite{Kriesten:2020apm} .  
Similar to the DVCS contribution, 
we can separate out the electric, magnetic and axial contributions, while simultaneously 
carrying out an analysis of the GPD content of the twist-two and twist-three parts to the cross section,
\begin{eqnarray}
\label{eq:siguuI}
{\cal I}   &= & e_l
 \, \frac{\Gamma}{ Q^2 \mid t \mid} \Re e \Big\{  A_{UU}^{\cal I}  \big(F_1 \mathcal{H} + \tau F_2  \mathcal{E} \big)   + B_{UU}^{\cal I}    G_M  \big( \mathcal{H}+ \mathcal{E} \big) 
 + C_{UU}^{\cal I} 
G_M \mathcal{ \widetilde{H} } + \frac{\sqrt{t_0-t}}{Q} \, F_{UU}^{{\cal I}, tw 3}  \Big\}  .  \nonumber \\
\end{eqnarray}
\begin{eqnarray}
\label{eq:Int_FUU3}
F_{UU}^{{\cal I}, tw 3} &= &  \Re e \Bigg\{ A^{ (3) \cal I}_{UU}  \Big[ F_{1}(2\mathcal{\widetilde{H}}_{2T} + \mathcal{E}_{2T}) + F_{2}( \mathcal{H}_{2T} + \tau \mathcal{\widetilde{H}}_{2T}) \Big]
+ B^{(3) \cal I}_{UU} G_M \, \widetilde{E}_{2T}  \nonumber \\
&+ & C^{(3) \cal I}_{UU}  G_M \, \Big[2\xi  H_{2T} - \tau( \widetilde{E}_{2T} -\xi E_{2T} )  \Big]
\Bigg\} \nonumber \\
&+&  \Re e \Bigg\{ \widetilde{A}^{(3) \cal I}_{UU}  \Big[F_{1}(2\widetilde{H}'_{2T} + E_{2T}') + F_{2}(H_{2T}' + \tau\widetilde{H}_{2T}' )\Big]  + \widetilde{B}^{(3) \cal I}_{UU}  G_M \widetilde{E}_{2T}' \nonumber \\
& + & 
  \widetilde{C}^{(3) \cal I}_{UU}  G_M \Big[2\xi  H_{2T}' - \tau( \widetilde{E}_{2T}' -\xi E_{2T}' )  \Big]\Bigg\} 
\end{eqnarray}

\section{Conclusions}
We presented a new approach and initial results from a collaborative effort including experts from theoretical physics and computer science, the EXCLAIM collaboration, aimed at obtain new physical information on the spatial structure of the proton and atomic nuclei from exclusive experiments. We argued that extracting information from data requires new methodologies and frameworks merging AI and theoretical physics ideas in a novel way. We are at a stage in our community where different efforts need to be benchmarked and coordinated. We proposed a set of such benchmarks \cite{Almaeen2021}.

On the other hand, in order to access the proton 3D structure we need to extend the number and  type of deeply virtual exclusive reactions with multiple particles in the final state.
It is therefore, important to write the cross sections for DVES which with a clear framework allowing for a QCD description where twist-two and twist-three effects are clearly demarcated.
Writing the cross section in terms of physically meaningful terms, {\it i.e.} underlying the we can understand more, and perform precise extractions as compared to a simple mathematical framework based on Fourier harmonics. 

\acknowledgments
This research is funded by DOE grants DE-SC0016286 and DE-SC0024644 (EXCLAIM collaboration).

\bibliographystyle{unsrt}
\bibliography{Biblio}

\end{document}